\documentstyle[11pt,epsfig,here,mine]{article}
\begin{document}

\begin{center}
{\Large
{\it
Submitted to the Proceedings of 25th ICRC\\
(Durban, South Africa, July 28 - August 8, 1997)\\
}}
\end{center}

\vspace{-1.4cm}
\title{RECONSTRUCTION OF ATMOSPHERIC NEUTRINOS   
WITH THE BAIKAL NEUTRINO TELESCOPE NT-96}
\vspace{0mm}

\vspace{14mm}
\author{
{\large THE BAIKAL COLLABORATION:}\\[2mm]
V.A.BALKANOV$^2$, I.A.BELOLAPTIKOV$^7$, L.B.BEZRUKOV$^1$, B.A.BORISOVETS$^1$, 
N.M.BUDNEV$^2$, A.G.CHENSKY$^2$, I.A.DANILCHENKO$^1$, ZH.-A.M.DJILKIBAEV$^1$, 
V.I.DOBRYNIN$^2$, G.V.DOMOGATSKY$^1$, A.A.DOROSHENKO$^1$, S.V.FIALKOVSKY$^4$,
O.N.GAPONENKO$^2$, A.A.GARUS$^1$, S.B.IGNAT$'$EV$^3$, A.KARLE$^8$, 
A.M.KLABUKOV$^1$, A.I.KLIMOV$^6$, S.I.KLIMUSHIN$^1$, A.P.KOSHECHKIN$^1$, 
V.F.KULEPOV$^4$, L.A.KUZMICHEV$^3$, B.K.LUBSANDORZHIEV$^1$, T.MIKOLAJSKI$^8$,
M.B.MILENIN$^4$, R.R.MIRGAZOV$^2$, A.V.MOROZ$^2$, N.I.MOSEIKO$^3$, 
S.A.NIKIFOROV$^2$, E.A.OSIPOVA$^3$, A.I.PANFILOV$^1$, YU.V.PARFENOV$^2$, 
A.A.PAVLOV$^2$, D.P.PETUKHOV$^1$, P.G.POKHIL$^1$, P.A.POKOLEV$^2$, 
M.I.ROZANOV$^5$, V.YU.RUBZOV$^2$, I.A.SOKALSKI$^1$, CH.SPIERING$^8$, 
O.STREICHER$^8$, B.A.TARASHANSKY$^2$, T.THON$^8$, D.B.VOLKOV$^2$, 
CH.WIEBUSCH$^8$, R.WISCHNEWSKI$^8$\\[2mm]
}

\address{
1 - Institute  for  Nuclear  Research,  Russian  Academy  of   Sciences
(Moscow); \mbox{2 - Irkutsk} State University (Irkutsk); \mbox{3 - Moscow}
State University (Moscow); \mbox{4 - Nizhni}  Novgorod  State  Technical
University  (Nizhni   Novgorod); \mbox{5 - St.Petersburg} State  Marine
Technical  University  (St.Petersburg); \mbox{6 - Kurchatov} Institute
(Moscow); \mbox{7 - Joint} Institute for Nuclear Research (Dubna);
\mbox{8 - DESY} Institute for High Energy Physics (Zeuthen) 
}

\vspace{1cm}
\vspace{-12pt}
\maketitle\abstracts{    
We describe the track reconstruction procedure
for events recorded with the  Neutrino Telescope {\it NT-96}.
After having identified 2 neutrino candidates close to the
opposite zenith with the small prototype telescope
{\it NT-36}, we present here results of the 
reconstruction of $5.3 \cdot 10^6$
muons, recorded with {\it NT-96} during its first 18 days lifetime.
We have separated  3 neutrino candidates, compared to 2.3 events
expected from MC calculations. 
}
\vspace{-6mm}
\section{Introduction}
   A primary challenge for deep underwater detectors is the
 identification of upward muons generated in  neutrino interactions. 
 Taking into account that the flux
 of downward muons at 1 km depth is about 6 orders of magnitude 
  larger than the flux  of upward muons, this task is extremely
  difficult with small detectors.
        
Data obtained with the 36-PMT array {\it NT-36}
showed good agreement of the observed zenith
angle distribution with that expected for
atmospheric muons \cite{JanICRC}. Also, the rate of fake events
(i.e. downward muons being reconstructed as upward muons)
was similar to MC calculations: 
S/N $\approx$ 1/40 for both
experimental data and the MC,
with $N$ being the fake rate, and $S$ the rate for events from true
upward  moving muons.
The agreement between {\it NT-36} data and MC gave confidence
in extrapolation of the results towards larger arrays.
However, with $S/N$ = 1/40, clear neutrino signatures were 
out of the range of {\it  NT-36} -- apart from a narrow cone 
around the opposite zenith \cite{APP}.

In 1996, the four-string array {\it NT-96} with 24 optical modules (OMs)
at each string was deployed.
With 72 m height, this detector surpassed
{\it NT-36} considerably, as well in the number of OMs as in 
the lever arm for track fitting.
The
 OMs were grouped in pairs along the string, with the
two PMTs switched in coincidence.
The orientation of the OMs was changed compared to the
symmetric {\it NT-36}. Only 
two of the twelve layers of {\it NT-96} (the second and  the eleventh) 
had upward  looking OMs, all others pointed down.
The strings
were placed at the edges of an trapezoid with side lengths of 3 $\cdot$
18.5 m, and 10.2 m \cite{status}.

\vspace{-2mm}
\section{Reconstruction procedure}
\vspace{-1mm}

The reconstruction algorithm is based on the assumption that the
light radiated by the muons is emitted exactly under the  Cherenkov
angle (42 degrees) with respect to the muon path. This 
"naked muon model" is a simplification,
since the direction of shower particles accompanying the muons 
is smeared around the muon direction. 
The reconstruction procedure consists of the following steps:
\begin{enumerate}
\vspace{-3mm}
\item A first quality analysis of the event which {\it a)}
   excludes events 
   far from being described by the model of a "naked muon", 
   and {\it b)} finds a first guess for the $\chi^2$ minimization.
\vspace{-3mm}
\item Determination of the muon trajectory based on the 
   minimization of the function
\begin{equation}
\chi^2_t = \sum_{i=1}^{N_{hit}} (T_i(\theta, \phi, u_0, v_0, t_0)
    - t_i)^2 / \sigma_{ti}^2
\end{equation}

\vspace{-3mm}
Here, $t_i$ are the measured times and $T_i$ the times expected for
a given set of track parameters. $N_{hit}$ is the number
of hit channels, $\sigma_{ti}$ are the timing errors. A set of
parameters defining a straight track is given by
$\theta$ and $\phi$ -- zenith and azimuth angle of the track,
respectively, $u_0$ and $v_0$ -- the two coordinates of
the track point closest to the center of the detector,
and $t_0$ -- the time the muon passes this point. 

\vspace{-3mm}
\item  Rejection of most bad reconstructed events 
  with the help  of final quality  criteria.
\end{enumerate}

\vspace{-3mm}
In the initial quality analysis (step 1),
an event has to pass the following criteria:
\begin{itemize}
\vspace{-3mm}
\item[a)] The time difference $\Delta t_{ij}$ must obey the 
	following condition:
\mbox{$|\Delta t_{ij}|\cos\eta<R_{ij}/c+\delta$}, here are $\eta$ the 
	Cherenkov angle and $R_{ij}$ the distance between
	channels. $\delta=$ 5ns. 
\vspace{-3mm}
\item[b)] For any two channels on the same string, a
   zenith angles region ${\theta^{min} -\theta^{max}}$ is determined which is allowed
   by the observed time differences:
\vspace{-3mm}
\begin{equation}
\label{eq:thetalimit}
\cos(\theta^{min}+\eta) < \cos\theta \frac{c \cdot \Delta t_{ij}}{z_j-z_i} < \cos({\theta^{max}-\eta})
\end{equation}

\vspace{-3mm}
  Here ${z_i,z_j}$ are z coordinates of the channels.  
   If the regions of possible zenith angles for all pairs
   along a string do not overlap, the event is
   excluded.
\vspace{-3mm}
\item[c)] Assuming the event is caused by a naked muon, 
   for every channel one can define a range of distances to
   the muon  depending on the measured amplitude $A$ of the
   channel. For every channel {\it pair} one can define the
   minimal ($\Delta t^{min}$) and maximal ($\Delta t^{max}$) allowed 
time difference , in dependence on the
   distance between the channels and the amplitudes.
If for any pair the condition
$\Delta t^{min} <\Delta t^{exp}<\Delta t^{max}$
is violated, the event is rejected.
\end{itemize}

\vspace{-3mm}
Seventy percent of the triggered events (trigger {\it 6/3},
meaning at least 6 hits at 3 strings.)
pass these criteria in both the experimental and the MC sample.
In the case of MC generated neutrino induced muons, the rate is
larger (80\%) due to the absence of muon bundles.

We apply {\it final} quality
cuts after the minimization (see item 3 above). For {\it NT-96} the most 
effective cuts are the traditional $\chi^2$ cut,
cuts on the probability of non-fired channels not to be hit,
and fired channels to be hit ($P_{nohit}$ and $P_{hit}$, respectively),
cuts on the correlation function of measured amplitudes
to the amplitudes expected for the reconstructed tracks,
and a cut on the amplitude $\chi^2$ defined similar to the
time $\chi^2$ defined above.

To guarantee a minimum lever arm for track fitting,
we reject events with a projection of the most distant 
channels on the track ($Z_{dist}$) below 35 meters. 
Due to the small transversal dimensions of {\it NT-96},
this cut excludes zenith angles close to the horizon, i.e.,
the effective area of the detector
with respect to atmospheric neutrinos is decreased considerably
(fig.1).

\vspace{-7mm}
\begin{figure}[t]
\centering
\mbox{\epsfig{file=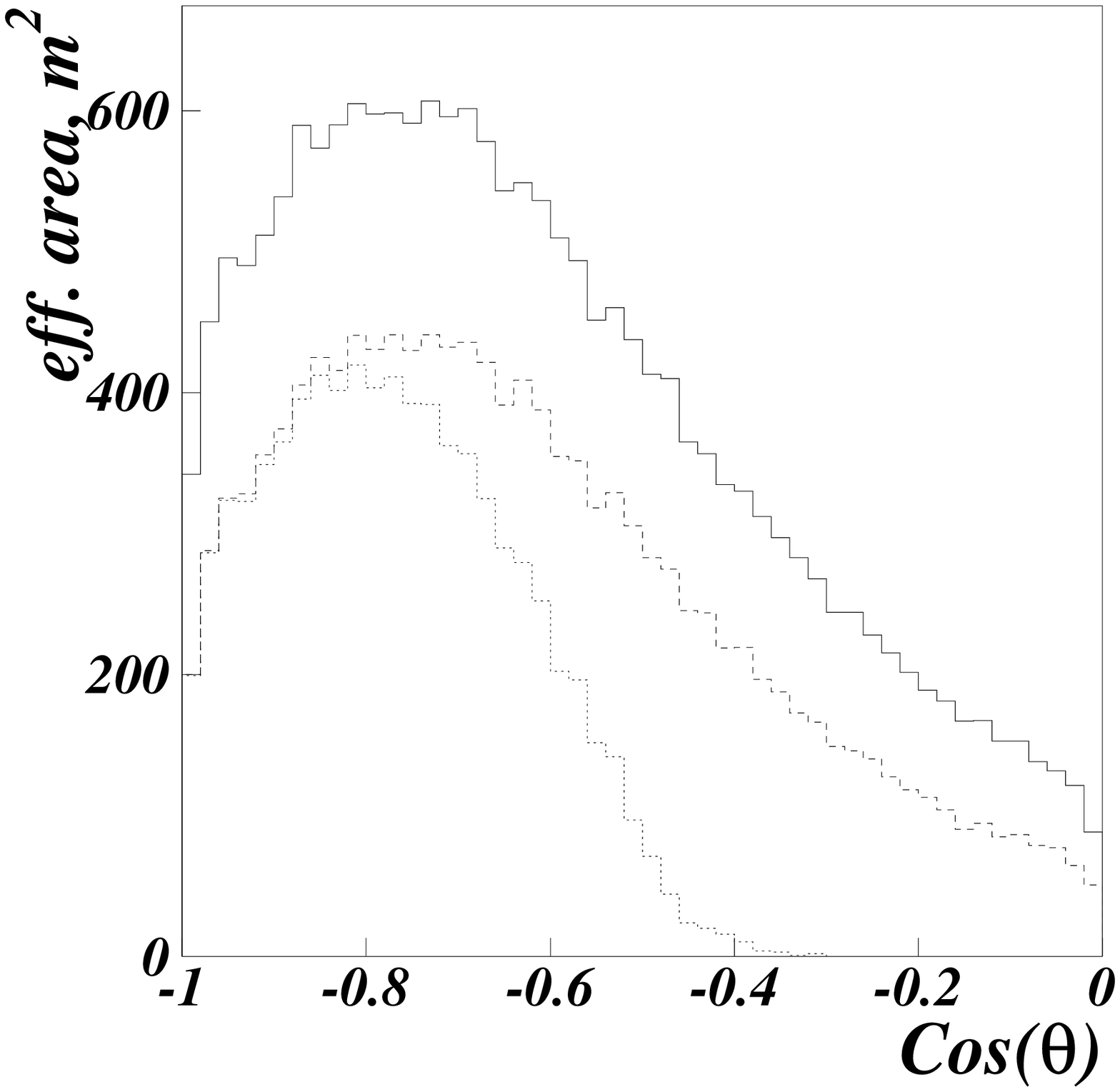,width=7.5cm,height=8.5cm}}
\mbox{\epsfig{file=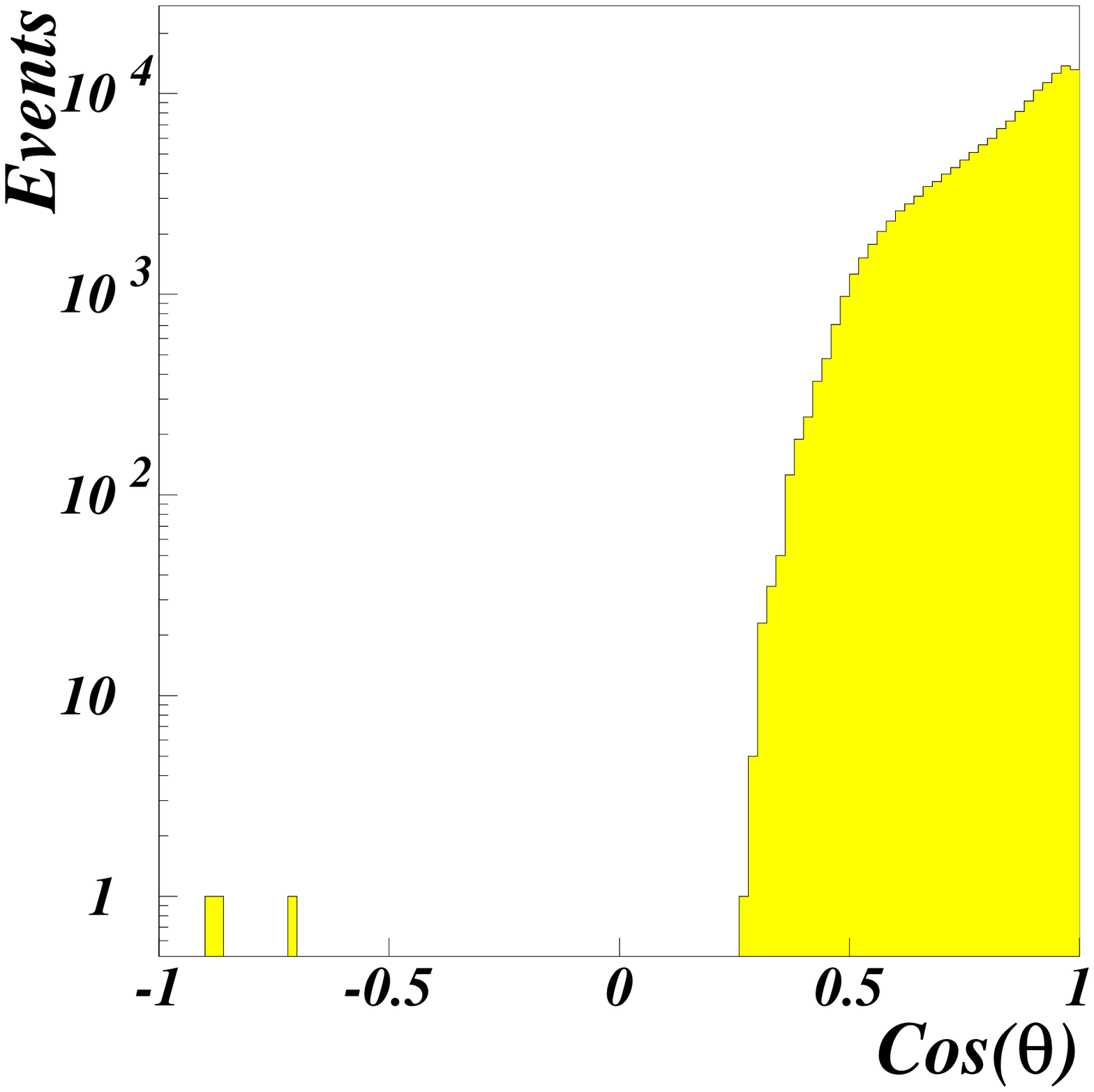,width=7.5cm,height=8.5cm}}
     \parbox[t]{7.2cm}{
          \caption [1] {
\frenchspacing
Effective area for upward muons satisfying trigger {\it 9/3};
solid line - no quality cuts; dashed line~-~final quality cuts;
dotted line - final quality cuts and restriction on $Z_{dist}$ (see text).
\nonfrenchspacing
}}
\hspace{.4cm}
     \parbox[t]{7.2cm}{\caption [2]{
\frenchspacing
Experimental angular distribution of events satisfying trigger 
{\it 9/3}, all final quality cuts and the limit on $Z_{dist}$ (see text).
\nonfrenchspacing
}}
\end{figure}

\vspace{3mm}
\section{Results}

The efficiency of all criteria was tested using MC generated atmospheric
muons and upward muons due to atmospheric neutrinos. 
$ 1.8 \cdot 10^6$ events from
atmospheric muon events (trigger {\it 6/3}) have been simulated, with
only 2 of them
passing all cuts and being reconstructed as upward going muons.
This corresponds to $S/N \approx 1$. 
Rejecting all events with less than 9 hits, no MC fake event
is left, with only a small decrease in neutrino sensitivity (see
table). This corresponds to $S/N > 1$ and the lowest curve
in fig.1.

Table 1
shows the fraction of events after the final quality
criteria, normalized to the number of events surviving pre-criteria
and reconstruction, for triggers {\it 6/3} and {\it 9/3}, respectively.

\vspace{6.5mm}
\makebox[49mm] {\scriptsize Table 1}

\vspace{0mm}
\begin{center}
\begin{tabular} {|c|c|c|c|} \hline
\raisebox{0pt}{Trigger cond.}&
\raisebox{0pt}{Experiment}&
\raisebox{0pt}{MC atm $\mu$}&
\raisebox{0pt}{MC $\mu$ from $\nu$} \\ \hline \hline
6/3 &   0.19 &           0.21    &        0.20 \\ \hline
9/3 &   0.044&           0.056   &       0.175 \\ \hline
\end{tabular}
\end{center}
\vspace{8mm}

With this procedure, we have reconstructed $5.3 \cdot 10^6$ events
taken with {\it NT-96} in April/May 1996. 
The resulting angular distribution is presented in 
fig.2. Three events were recognized as upward going muons.
Fig.3 displays one of the neutrino candidates. 
Top right the times of the hit channels are shown as
as function of the vertical position of the channel.
At each string we observe the
time dependence characteristically for upward moving particles.

\clearpage

\begin{figure}[h]
\centering
\mbox{\epsfig{file=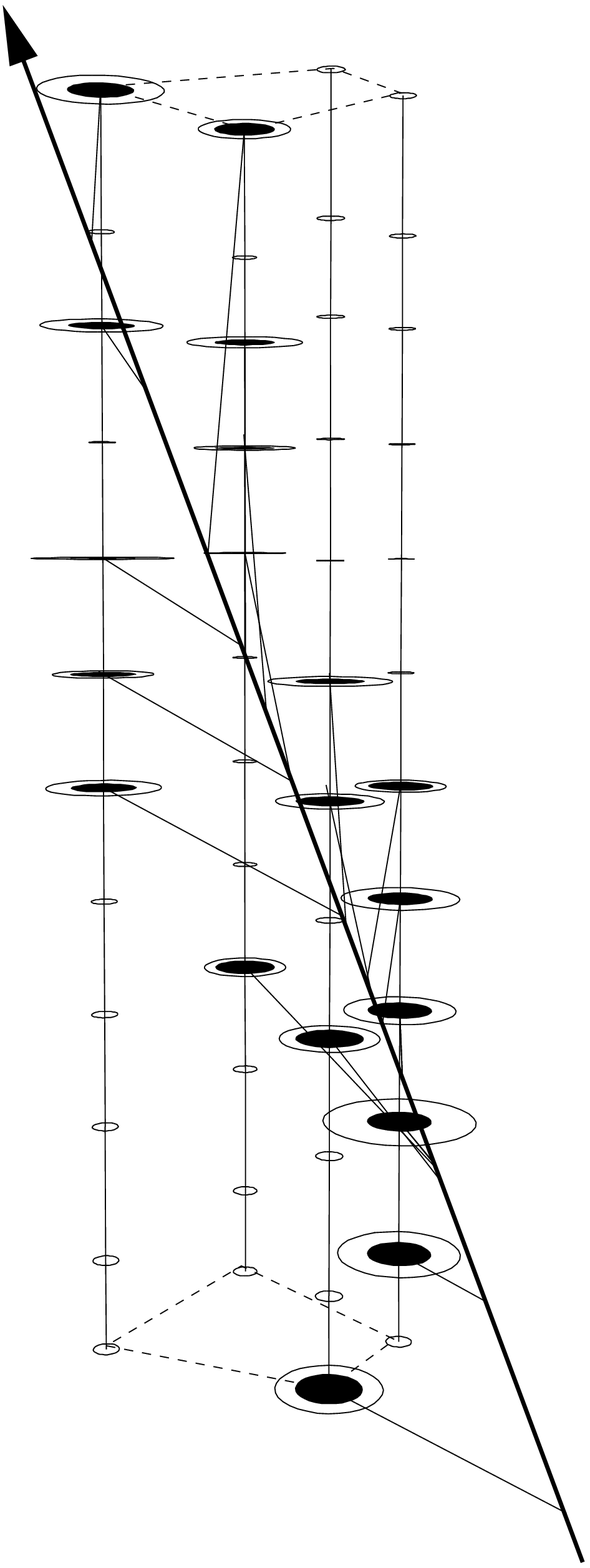,width=6cm,height=15cm}}
\mbox{\epsfig{file=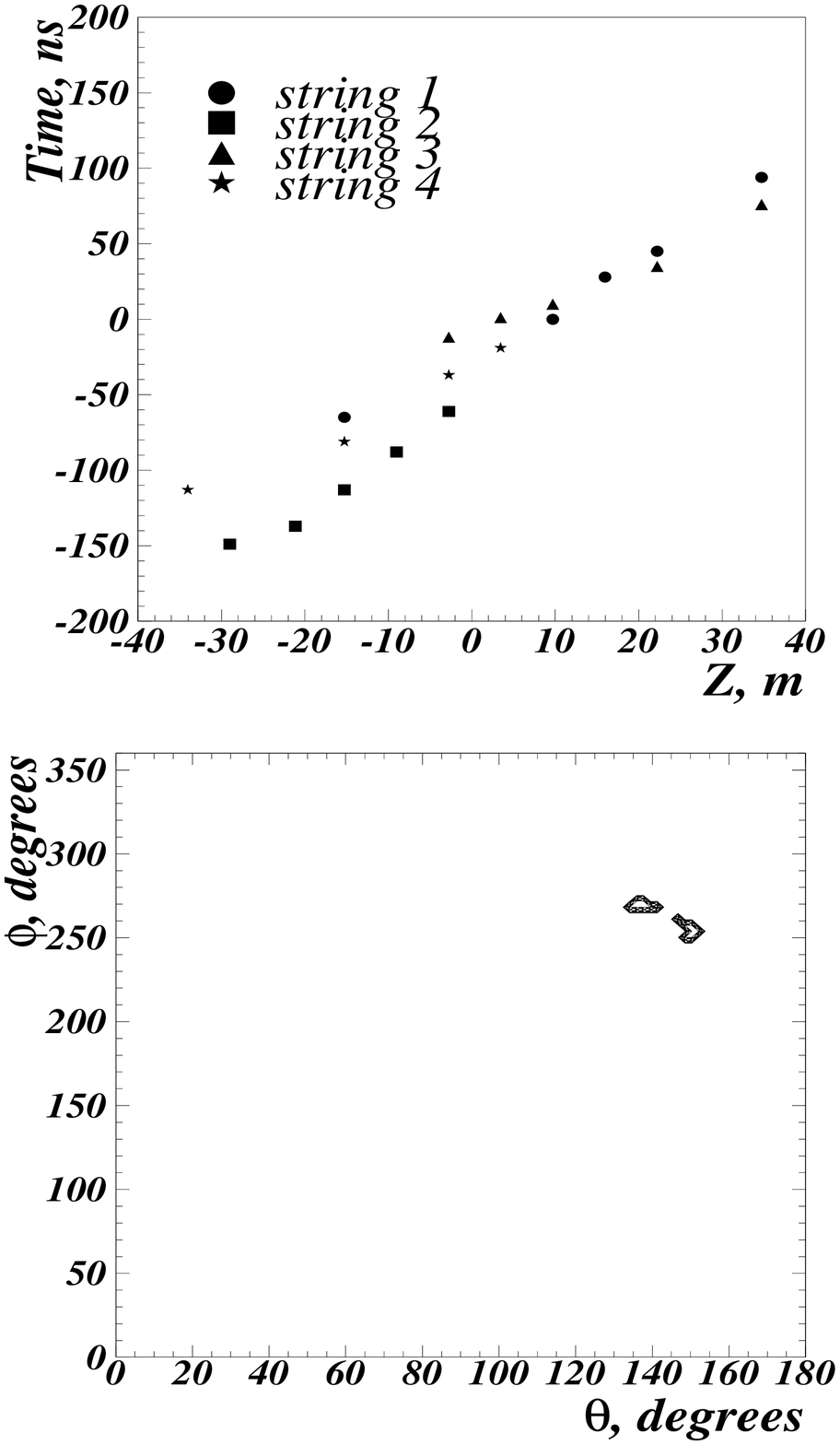,width=9cm,height=15cm}}
{\caption[3]{
\frenchspacing
A "gold plated" 19-hit neutrino event. {\it Left:} Event display.
Hit channels are in black. The thick line gives the 
reconstructed muon path, thin lines pointing to the channels mark the path
of the 
Cherenkov photons as given by the fit to the measured times. The sizes of 
the ellipses are proportional to the recorded amplitudes. {\it Top right:} 
Hit times versus vertical channel positions. 
{\it Bottom right:}  The allowed $\theta/\phi$ regions (see text).
\vspace{6mm}
\nonfrenchspacing
}}
\end{figure}

\vspace{-8mm} 
Applying eq.\ref{eq:thetalimit} 
not only to pairs at the same string, but
to all pairs of 
hit channels, one can construct an allowed region in both 
$\theta$ and $\phi$. 
For clear neutrino events this region is situated totally below horizon.
This is demonstrated at the bottom right picture of fig.3.
The same holds for the other two events, one of which is shown
in fig.4. Fig.5, in contrast, shows an ambiguous event giving,
apart from the upward solution, also a downward solution. In this case
we assign the event to the downward sample.

\nopagebreak

\section{Conclusions}

The analysis presented here is based on the data taken with {\it NT-96}
between April 16 and May 17, 1996 (18 days lifetime). Three
 neutrino candidates
have been separated, in good 
agreement with the expected 
number of upward events of approximately 2.3. Our 
algorithm\linebreak 

\begin{figure}[h]
\centering
\mbox{\epsfig{file=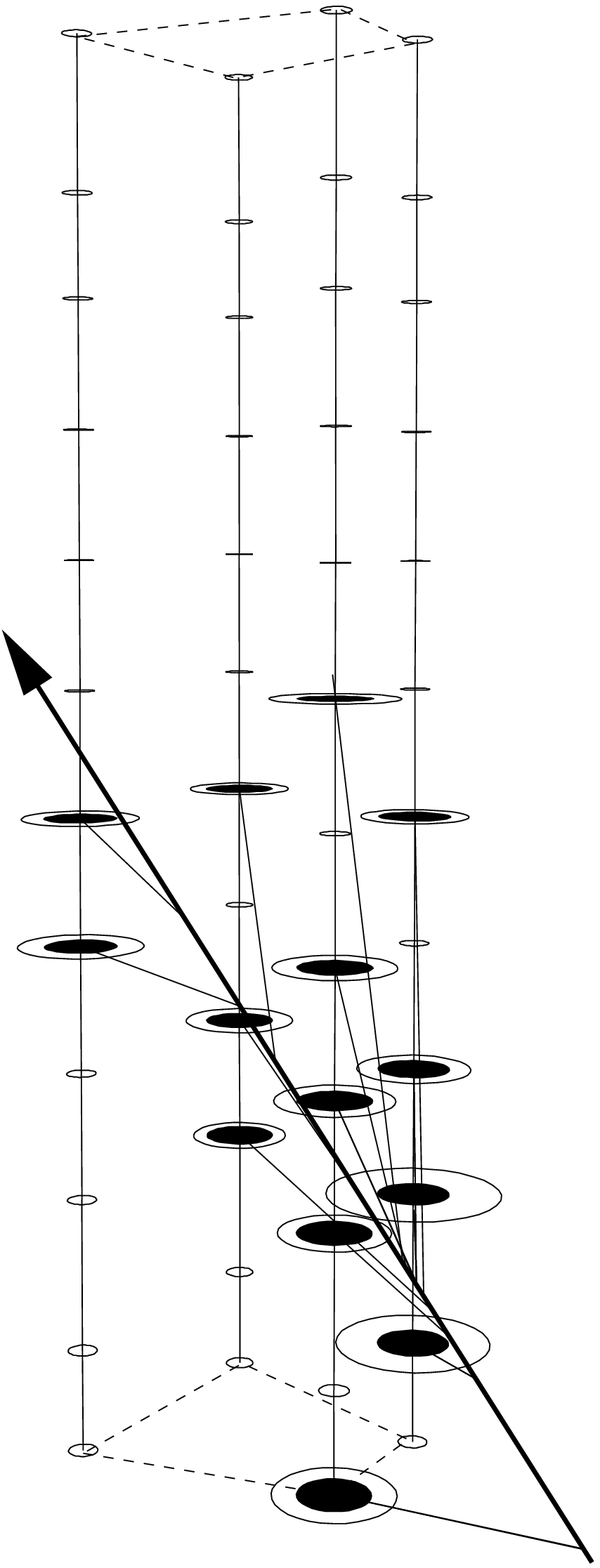,width=7cm,height=12.7cm}}
\mbox{\epsfig{file=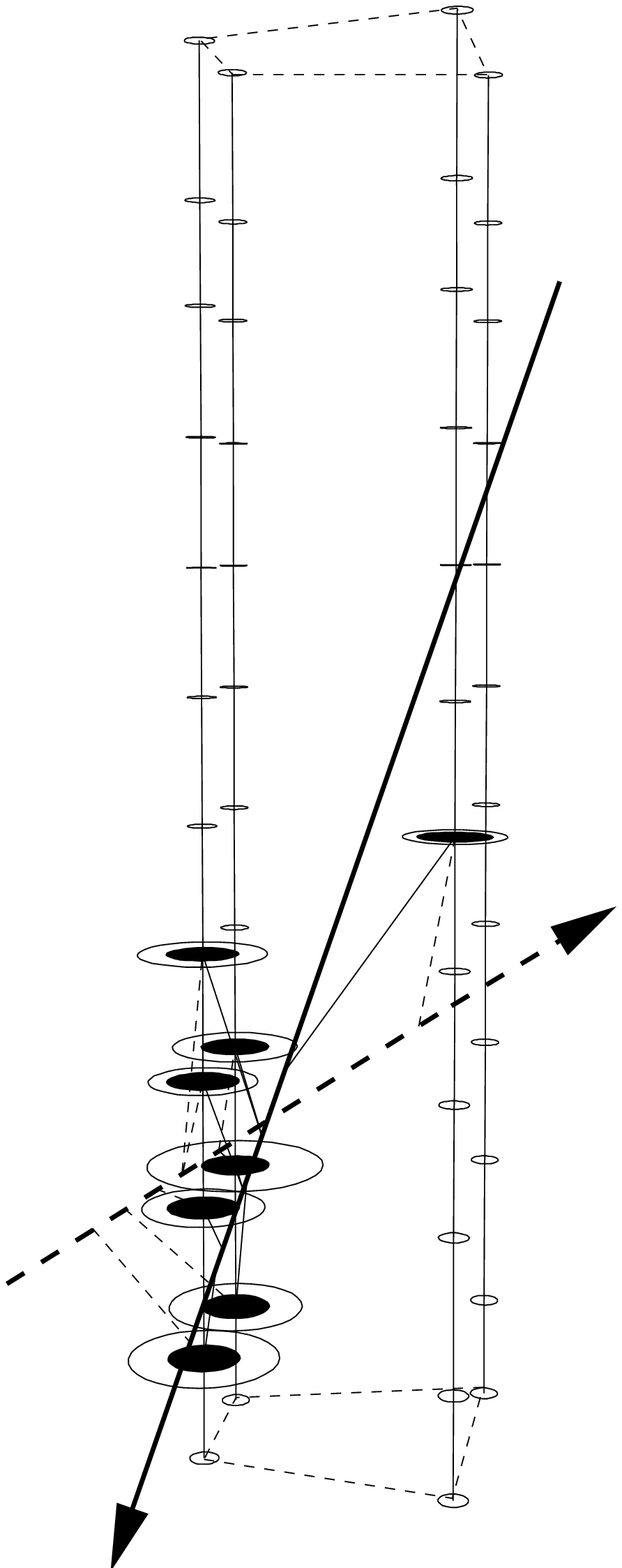,width=7cm,height=12.7cm}}
     \parbox[t]{7cm}{
          \caption [4] {
\frenchspacing
An unambiguous 14-hit neutrino candidate.
\nonfrenchspacing
}}
\hspace{.4cm}
     \parbox[t]{7cm}{\caption [5]{
\frenchspacing
An ambiguous event reconstructed as a neutrino event (dashed line) but with
a second solution above the horizon (solid line).
This event was assigned to the sample of downward going muons.
\nonfrenchspacing
}}
\end{figure}

\noindent
allows to select neutrino
events in a cone with about 50 degrees half-aperture around 
the opposite zenith, and an effective area of $\sim 350 m^2$. 

With the experimental confirmation that {\it NT-96} can operate
as a neutrino detector, we now are searching for 
additional possibilities to reject fake events with a smaller loss 
in effective area. The increased transversal
dimensions of the future {\it NT-200} (1998) will significantly 
increase effective area and angular acceptance for reliably 
separatable up-going events.

\end{document}